\documentclass[aps,prd,twocolumn,amsmath,floatfix,longbibliography]{revtex4-1}
\usepackage{lineno,isotope,mathrsfs}
\modulolinenumbers[5]
\usepackage{amsmath,bbold,amssymb,epsfig,bm,mathrsfs,feynmp}
\usepackage{color,slashed,nicefrac,exscale,multirow,soul}
\usepackage{times,txfonts,xcolor,graphicx,gensymb,tikz}
\usepackage[normalem]{ulem}
\usepackage[breaklinks,colorlinks,citecolor=blue]{hyperref}
\definecolor{red}{rgb}{0.8,0,0}
\definecolor{violet}{rgb}{0.4,0,0.4}
\definecolor{green}{rgb}{0,0.5,0.0}
\definecolor{navy}{rgb}{0.0,0.0,0.6}
\definecolor{orange}{rgb}{0.8,0.2,0.0}

\usepackage[normalem]{ulem}  

\newcommand{\bea}{\begin{eqnarray}}
\newcommand{\eea}{\end{eqnarray}}
\newcommand{\ep}{\varepsilon}
\newcommand{\MR}{$M$-$R\ $}

\newcommand{\Lsym}{\ensuremath{L_{\text{sym}}}}
\newcommand{\Qsat}{\ensuremath{Q_{\text{sat}}}}
\newcommand{\rhotran}{\ensuremath{\rho_{\text{tran}}}}

\begin{document}
\title{Relativistic hybrid stars in light of the NICER
PSR J0740+6620 radius measurement}
\author{Jia Jie Li}
\email{jiajieli@swu.edu.cn}
\affiliation{School of Physical Science and Technology, Southwest University, Chongqing 400700, China}
\author{Armen Sedrakian}
\email{sedrakian@fias.uni-frankfurt.de}
\affiliation{Frankfurt Institute for Advanced Studies,
D-60438 Frankfurt am Main, Germany}
\affiliation{Institute of Theoretical Physics,
University of Wroclaw, 50-204 Wroclaw, Poland}
\author{Mark Alford}
\email{alford@physics.wustl.edu}
\affiliation{Department of Physics, Washington University,
St.~Louis, Missouri 63130, USA}

\begin{abstract}
We explore the implications of the recent radius determination of
PSR J0740+6620 by the NICER experiment combined with the neutron
skin measurement by the PREX-II experiment and the associated inference of the slope of symmetry energy, for the structure of hybrid
stars with a strong first-order phase transition from nucleonic to
quark matter. We combine a covariant density-functional nucleonic 
equation of state (EOS) with a constant-speed-of-sound EOS for
quark matter. We show that the radius and tidal deformability ranges
obtained from GW170817 can be reconciled with the implication of the PREX-II experiment
if there is a phase transition to quark matter in the low-mass 
compact star. In the high-mass segment, the EoS needs to be stiff 
to comply with the large-radius inference for PSR J0740+6620
and J0030+0451 with masses $M\simeq 2M_{\odot}$ and
$M\simeq 1.4M_{\odot}$. We show that twin stars are not excluded, but the mass and radius ranges (with $M \geq M_\odot$) are restricted to narrow domains
$\Delta M_{\rm twin} \lesssim 0.05 M_\odot$ and
$\Delta R_{\rm twin} \sim 1.0$~km. We also show that the existence 
of twin configurations is compatible with the light companion in the
GW190814 event being a hybrid star in the case of values of the
sound-speed square $s = 0.6$ and  $s=1/3$.
\end{abstract}
\maketitle
%
\section{Introduction}
\label{sec:Intro}

The recent mass measurement of the heaviest pulsar to date 
PSR J0740+6620~\citep{NANOGrav_2019}, which is currently at 
$2.08^{+0.07}_{-0.07}\,M_\odot$ and the recent analysis~\citep{NICER_2021a,NICER_2021b}
of x-ray data from the NICER experiment, which gave 
the radius estimates
$12.39^{+1.30}_{-0.98}$~\citep{NICER_2021a} and
$13.71^{+2.61}_{-1.50}$ km~\citep{NICER_2021b} and corresponding 
mass estimates $2.07^{+0.07}_{-0.07}\,M_{\odot}$ and
$2.08^{+0.09}_{-0.09}\,M_{\odot}$ of PSR J0740+6620
open prospects of constraining 
the properties of dense-matter equation of state (EOS)---in particular, the possibility of a phase transition at high densities.

The lower bound on the radius of PSR J0740+6620 constrains a highly
relevant region of the mass-radius ($M$-$R$) diagram of compact stars
(CSs), which in combination with the determination of the tidal
deformability (TD) of a star of mass $\sim 1.4\,M_\odot$ in the GW170817
event by the LIGO-Virgo Collaboration~\citep{LIGO_Virgo_2019}, puts
significant constraints on the EOSs of CSs. These require that the
stellar EOS must be moderately soft at intermediate densities (to
allow for relatively small TDs) and must be stiff enough at high
densities (to allow for two-solar-mass CSs; see Refs.
\citep{Tanhuang_2021,Biswas_2021,Legred_2021,Raaijmakers_2021,Huth_2021,Zhangnb_2021,Tangsp_2021}).

Recently, the Lead Radius Experiment Collaboration (PREX-II)
  reported the most precise measurement yet of the neutron skin
  thickness of the lead nucleus 
  $R^{208}_{\rm skin}= 0.283 \pm 0.071$~fm (mean and $1\,\sigma$
  standard deviation), in a
  parity-violating electron scattering experiment~\citep{PREX-II_2021}. Subsequent theoretical
  analysis~\cite{Reed_2021,Reinhard_2021} based on density-functional
  theory established values of the symmetry energy $E_{\rm sym}$ and
  the slope of nuclear symmetric energy $L_{\rm sym}$ at saturation
  density ($\rho_{\rm sat}$) that are consistent with the inferred
  value of $R^{208}_{\rm skin}$. Reference~\cite{Reed_2021} finds
  $E_{\rm sym} = 38.1 \pm 4.7$~MeV and
  $L_{\rm sym} = 106 \pm 37$~MeV from a family of relativistic
  (nonlinear, meson-exchange) density functionals
  (DFs). Reference~\cite{Reinhard_2021} expanded the base (and the functional form) of employed DFs to include nonrelativistic DFs,
  relativistic DFs 
  with density-dependent meson-exchange couplings and
  relativistic point coupling DFs 
  to find $E_{\rm sym} = 32 \pm 1$
  MeV and $L_{\rm sym} = 54 \pm 8$ MeV. These 
  values 
  include the additional requirement on DFs to be consistent 
  with the 
  experimental limits on the dipole polarizability of $^{208}$Pb, which prefer DFs predicting a small 
  value of $L_{\rm sym}$. The
  large value of $L_{\rm sym}$ found in the first analysis is in potential
  tension with the various
  estimates~\cite{Lattimer2013ApJ,Danielewicz2014NuPhA,Oertel_2017,BaldoBurgio_2016},
  whereas the value obtained from the second analysis is within the
  range inferred previously. Note that the difference between the
  two quoted values of $L_{\rm sym}$ is $1.37\,\sigma$ which translates
  to about $83\%$ significance. Since $L_{\rm sym}$ is highly correlated with
  the stellar radius and TD, the rather large value
  found in the first analysis is in potential tension with the
  GW170817 deformability measurement if one assumes a purely nucleonic
  composition~\citep{Reed_2021}.

Nevertheless, independent of the interpretation of the PREX-II experiment, 
the potential tension outlined above suggests a close examination of
the compatibility of a large value of $L_{\rm sym}$ with the constraints
on compact stars. Here we address the possibility of 
{\it first-order phase transition} from nucleonic to quark matter
using (a)~the radius measurements of two neutron stars by NICER, (b)~the TD of GW170817, and (c)~a range of values of $L_{\rm sym}$ suggested by Refs.~\cite{Reed_2021,Reinhard_2021}.

The composition of matter at high densities achieved in a CS's core
remains unknown. One possibility is a deconfinement phase transition
from bound states of hadrons to liberated quark
states~\citep{Alford_2008}.
If the quark matter in a CS's core is relatively soft then its radius 
and TD are reduced, which could avoid the 
tentative tension between the inferred 
TDs from the GW170817 event and those predicted by purely hadronic stiff 
EOS models without a phase 
transition~\citep{Annala_2018,Paschalidis_2018,Most_2018,Tews_2018,Burgio_2018,Alvarez-Castillo_2019,Christian_2019,Montana_2019,Sieniawska_2019,Essick_2019,Lijj_2020a,Miaozq2020,Liang2021,Malfatti2020,Rodriguez2021}. Of particular interest in this scenario is the emergence of 
{\it twin and triplet stars}, where one (or two) hybrid stars have 
the same mass as, but a different radius from, a purely hadronic 
star~\citep{Alford_2013,Alford_2017,Paschalidis_2018,Alvarez-Castillo_2019,Christian_2019,Montana_2019,Lijj_2020a}. The more extreme case of triplets will not be discussed here.

With a mass of about $2.1\,M_\odot$, PSR J0740+5620 is the most massive
known CS: it is about $50\%$ more massive than PSR J0030+0451. 
Yet current measurements do not indicate a significant difference 
in size; see Fig.~\ref{fig:TempoN}. As mentioned earlier, the generic 
feature of the first-order phase transition is the softening of the 
EOS in the vicinity of the transition point, which leads to smaller 
radii of hybrid stars as compared to the nucleonic ones. Confronting 
the very massive star's radius estimation with the possible phase 
transition in dense matter is thus very timely. This issue has been 
discussed by Refs.~\cite{Legred_2021,Tanhuang_2021}.

In this work, we apply the setup developed by us earlier~\citep{Lijj_2020a} 
to explore the consequences of a strong first-order phase transition from 
nucleonic to quark matter and the formation of twin stars. The resulting 
models are then confronted with the newly available data from NICER and PREX-II experiments.

\section{Construction of EOS}
\label{sec:Construction}
%
\begin{figure}[tb]
\centering
\ifpdf
\includegraphics[width = 0.45\textwidth]{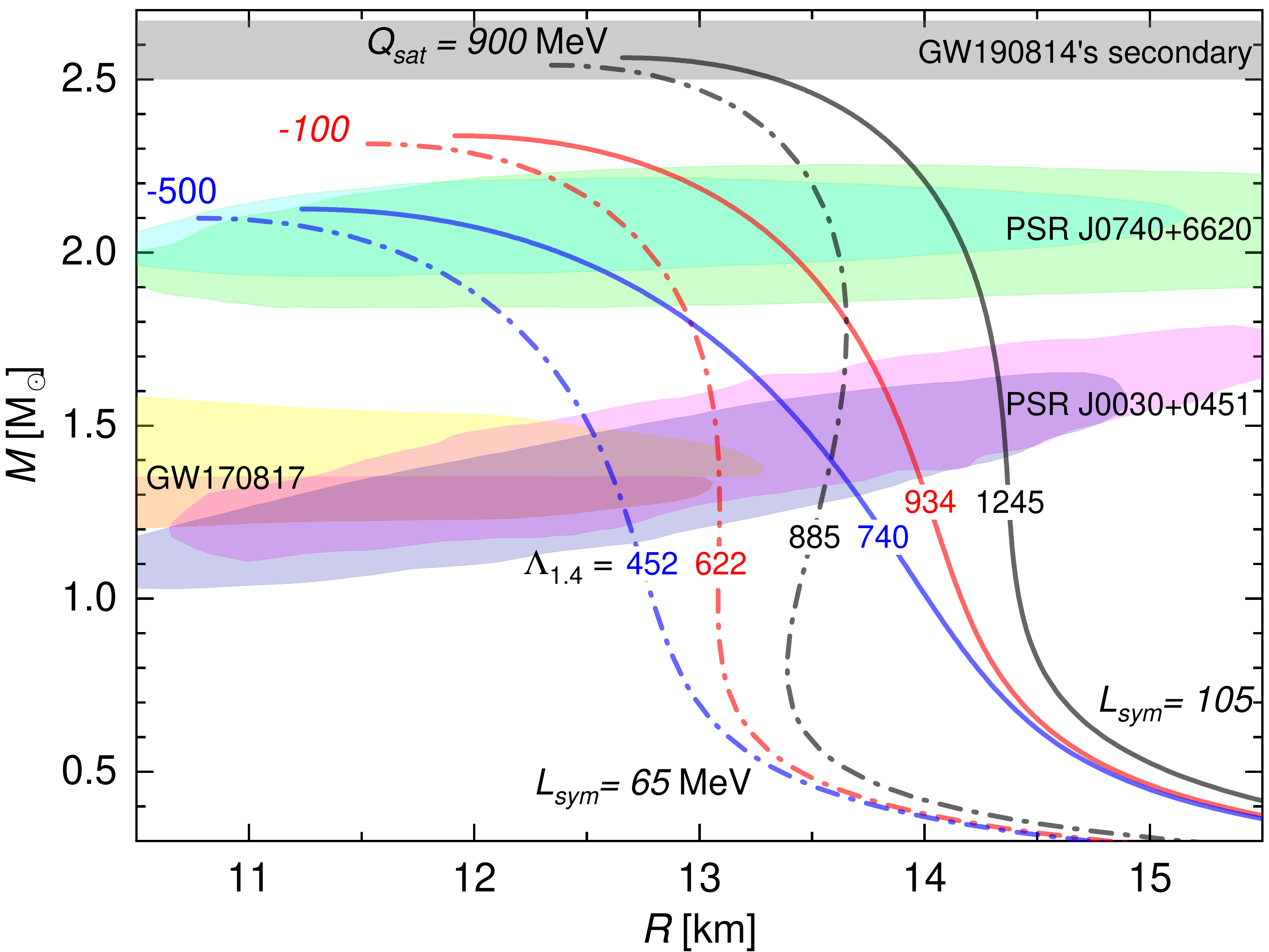}
\else
\includegraphics[width = 0.45\textwidth]{TempoN.eps}
\fi
\caption{\MR relation for nucleonic EOS for different pairs of values
of $Q_{\rm sat}$ and $L_{\rm sym}$. In addition we show the TD of a 
canonical CS for each model. Constraints with 90\% credibility from multimessenger 
astronomy are shown by shaded regions~\citep{LIGO_Virgo_2019,NICER_2019a,NICER_2021a,NICER_2019b,NICER_2021b}; see text for details.}
\label{fig:TempoN}
\end{figure}

To describe low-density matter, we use a covariant density functional (CDF) 
of nuclear matter in which baryons are coupled to mesons with density-dependent 
couplings, and we adopt as a reference the DDME2 parametrization~\citep{Lalazissis_2005}.  
The CDF EOS models can be extended by modifying the density dependence of the
couplings~\citep{Lijj_2019b} such as to map the CDF EOS models onto
purely phenomenological ones~\citep{Margueron_2018}, which are based on
an expansion of the energy density of nuclear matter close to the
saturation density $\rho_{\rm sat}$ with respect to the density and
isospin asymmetry---i.e.,
\begin{eqnarray}
\label{eq:Taylor_expansion}
  E(\chi, \delta) & \simeq & E_{\text{sat}} + \frac{1}{2!}K_{\text{sat}}\chi^2
                             + \frac{1}{3!}Q_{\text{sat}}\chi^3 \nonumber \\ [1.0ex]
                        &  & +\,E_{\text{sym}}\delta^2 + L_{\text{sym}}\delta^2\chi
                             + {\mathcal O}(\chi^4, \chi^2\delta^2),
\end{eqnarray}
where $\chi=(\rho-\rho_{\text{sat}})/3\rho_{\text{sat}}$ with $\rho$ 
and $\rho_{\text{sat}}$ being the  number density and its value at saturation, and
$\delta = (\rho_{n}-\rho_{p})/\rho$ where $\rho_{n(p)}$ is
the neutron(proton) number density. The coefficients of the expansion
are known as the {\it incompressibility} $K_{\text{sat}}$, the 
{\it skewness} $Q_{\text{sat}}$, the {\it symmetry energy}
$E_{\text{sym}}$, and the {\it slope parameter} $L_{\text{sym}}$. 
In this manner, the uncertainties in the gross properties of CSs
can be quantified in terms of not-well-known higher-order 
{\it characteristics of nuclear matter}, specifically $Q_{\rm sat}$ 
and $L_{\rm sym}$~\citep{Lijj_2019b,Margueron_2018}. The low-order
characteristics are fixed at their values predicted by the DDME2 
parametrization: $E_{\rm sat} = -16.14$, $K_{\rm sat} = 251.15$,
$E_\text{sym} = 32.31$\,MeV and $\rho_{\rm sat} = 0.152$ fm$^{-3}$. 
At the same time, the CDF provides access to the composition of matter which
is not fixed in agnostic models based on nuclear characteristics only.
 
Motivated by this, we construct six representative EOSs featuring
combinations of three values of
$Q_{\rm sat} = -500$, -100, 900 MeV and two values of
$L_{\rm sym} = 65$, 105 MeV. The value of $Q_{\rm sat}$ controls the
high-density behavior of the EOS, and thus, the maximum mass of a static 
CS~\citep{Lijj_2019b}. For $Q_{\rm sat} = -500$\,MeV the maximum mass is about 2\,$M_\odot$, 
which matches the mass measurement of PSR J0740+6620~\citep{NANOGrav_2019,Fonseca_2021}; 
for $Q_{\rm sat} = -100$\,MeV the maximum mass is consistent with the
(approximate) {\it upper limit} on the maximum mass of static CSs 
$\sim2.3\,M_\odot$ inferred from the analysis of GW170817 event~\citep{Rezzolla_2018,Khadkikar_2021};
finally, for $Q_{\rm sat} = 900$\,MeV the maximum mass is close to
2.5\,$M_\odot$, which would be compatible with the mass of the secondary
in the GW190814 event~\citep{LIGO-Virgo_2020} and its interpretation
as a nucleonic CS~\citep{Fattoyev_2020,Lijj_2020b}. We chooe values of 
$L_{\rm sym}$ corresponding to the central value and the lower range of
the $1\,\sigma$ confidence interval (CI) of the PREX-II measurement, 
$\Lsym = 106\pm 37$\,MeV
\citep{PREX-II_2021,Reed_2021}. Note that the recent
measurement of the spectra of pions in intermediate energy collisions
implies $L_{\rm sym} = 79.9\pm 37.6$~MeV~\citep{SRIT_2021}. In addition, an analysis based on nonparametric EoS shows that there is a mild tension between the results of Ref.~\cite{Reed_2021} and astrophysical data supplemented by chiral effective field theory results~\cite{Essick2021}.

The \MR relations for our six EOS are shown in Fig.~\ref{fig:TempoN},
where we also show the current astrophysical observational constraints. We show 90\%-credible ellipses from each of the two NICER modeling groups 
for PSR~J0030+0451 and J0740+6620~\citep{NICER_2019a,NICER_2021a,NICER_2019b,NICER_2021b}. 
We also show 90\%-credible regions for each of the two CSs that merged
in the gravitational wave event GW170817~\citep{LIGO_Virgo_2019}. 
Finally, we show the 90\% CI for the mass of the secondary component
of GW190814~\citep{LIGO-Virgo_2020}. 
A lower limit on the average TD $\tilde\Lambda_{1.186}\ge 240$ (with binary chirp mass $\mathcal{M} = 1.186\,M_\odot$) was extracted from the GW170817 event using the observations and modeling of this merger~\cite{Radice2018,Kiuch2019}. This limit allows for hybrid stars having $R_{1.4}$ radii~\cite{Burgio_2018} well below the  corresponding NICER
limits~\citep{NICER_2019a,NICER_2019b}. 

The softest EOS model ($Q_{\rm sat} = -500$, $L_{\rm sym} = 65$~MeV),
which gives $R_{1.4} = 12.58$~km and $\Lambda_{1.4}= 452$ for canonical-mass
star, appears as the only model satisfying all three of the
\MR constraints. The stiffest EOS model ($Q_{\rm sat} = 900$,
$L_{\rm sym} = 105$ MeV) which gives $R_{1.4} = 14.37$~km and
$\Lambda_{1.4} = 1245$, passes  through the upper range of the
CIs provided by NICER results. The $M$-$R$ relations shown in 
Fig.~\ref{fig:TempoN} indicate that one could find a compromise 
between the requirements of a soft EoS (from GW170817) and
relatively large value of $L_{\rm sym}$ implied by the analysis of 
PREX-II in Ref.~\cite{Reed_2021}.
This would be accomplished by an EOS with $L_{\rm sym}\sim 65$\,MeV, 
which is at the lower end of $1\,\sigma$ range of Ref.~\cite{Reed_2021}, and a 
negative $Q_{\rm sat}$.

So far, we have seen that fairly low values of $L_{\rm sym}$ from the
range consistent with the inference~\cite{Reed_2021}  from 
the PREX-II experiment allow us to build models
that are consistent with the known astronomical constraints. However,
as we show next, larger values from the experimental range of
$L_{\rm sym}$ can be accommodated if a strong first-order phase
transition to quark matter is allowed for. We will model below the
EOS of the quark phase using the constant-sound-speed (CSS)
parametrization~\citep{Zdunik_2013,Alford_2013}, which matches well
with the predictions based on the Nambu--Jona-Lasinio (NJL) model 
computations which include vector repulsion~\citep{Bonanno_2012,Blaschke_2010}.
We assume a first-order phase transition with a sharp boundary between 
the nucleonic and quark phases (which is the case when mixed phases 
are disfavored by surface tension and electrostatic energy
costs~\citep{Alford_2001}). The pressure is then given
by~\citep{Zdunik_2013,Alford_2013}
\bea\label{eq:EoS}
p(\ep) =\left\{
\begin{array}{ll}
p_{\rm tran}, \,\, & \ep_{\rm tran} < \ep < \ep_{\rm tran}\!+\!\Delta\ep, \\[0.5ex]
p_{\rm tran} + s\,\bigl[\ep-(\ep_{\rm tran}\!+\!\Delta\ep)\bigr],
  \,\, & \ep_{\rm tran}\!+\!\Delta\ep < \ep, 
\end{array}
\right.
\eea
where $p_{\rm tran}$ is the transitional pressure with energy density
$\ep_{\rm tran}$, $\Delta\ep$ is the discontinuity, and $s$ is the
square of the sound speed in the quark phase. The possible topologies
of hybrid stars in the \MR diagram based on CSS parametrization have
been studied~\citep{Alford_2013}. Of particular interest is the case
where, by an appropriate choice of the parameters $p_{\rm tran}$,
$\Delta\ep$, and $s$, there are two disconnected branches of stars: one
with purely nucleonic and the other with hybrid stars. Such topology
leads, in particular, to twin configurations where stars have
identical masses but different
radii~\citep{Paschalidis_2018,Alvarez-Castillo_2019,Christian_2019,Montana_2019}. 
The more extreme case of three disconnected branches leads to the 
formation of triplets, which will not be studied here~\citep{Alford_2017}.

\begin{figure}[tb]
\centering
\ifpdf
\includegraphics[width = 0.45\textwidth]{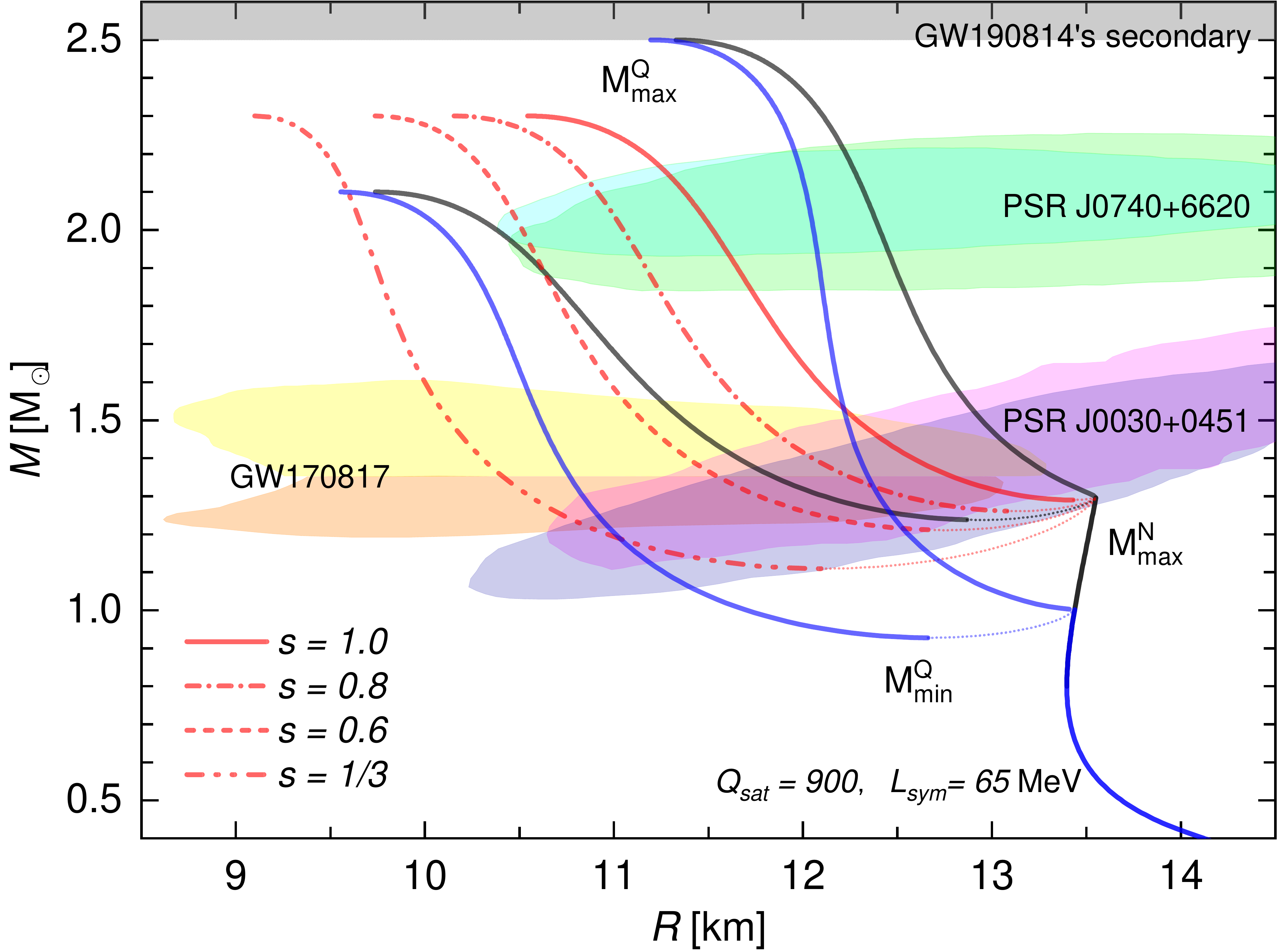}
\else
\includegraphics[width = 0.45\textwidth]{TempoH.eps}
\fi
\caption{
Illustrative \MR relation for hybrid EOS models. 
The results are constructed by varying one of the three quantities 
that fully determine the model at fixed values of the two others. 
These are the maximum masses of the quark 
$M^{\rm Q}_{\rm max}$ and nucleonic $M^{\rm N}_{\rm max}$ 
branches and the sound-speed square $s$. 
The dotted thin lines indicate unstable configurations.
All the solid curves correspond to $s=1.0$. For the remaining 
curves, the value of $s$ is as indicated in the plot.
}
\label{fig:TempoH}
\end{figure}

Figure~\ref{fig:TempoH} illustrates how such a transition to quark
matter allows a nuclear EOS with a large $\Lsym$ to be consistent with
astrophysical constraints. We have fixed the nuclear EOS by choosing
$\Lsym=65$ and $\Qsat=900$~MeV. We vary the CSS parameters of the
quark-matter EOS to explore different sound speeds $s$, and by varying
$\rhotran$ and $\Delta\ep_\text{tran}$ we can explore different values
of the maximum masses $M^{\rm N}_{\rm max}$ and $M^{\rm Q}_{\rm max}$ 
on the nucleonic branch and hybrid branch respectively.
In particular, each of the two same-color solid lines differ only by the value of $M^{\rm Q}_{\rm max}$.

The sequences shown in Fig.~\ref{fig:TempoH} are for
$M^{\rm Q}_{\rm max}/M_\odot= 2.1$, 2.3, 2.5 and
$M^{\rm N}_{\rm max}/M_\odot=1.0$, 1.3 (corresponding to transition
densities $\rhotran/\rho_{\rm sat} = 1.81$, 2.05).
We see that a stiff nucleonic EOS can be made compatible with the
GW170817 constraint if there is a first-order transition to quark
matter at densities $\rhotran\lesssim 2\,\rho_\text{sat}$.

Figure~\ref{fig:TempoH} shows, in addition, the sensitivity of results
toward varying the value of $s$ for a specific case with fixed
maximum masses. Since a reduction of $s$ from its maximal value
softens the quark-matter EOS, the \MR curves are shifted to the left,
eventually putting some of them ($s \lesssim 0.6$) outside of the NICER
result for PSR J0740+6620. Note that in Fig.~\ref{fig:TempoH} two
topologies of \MR curves are present: the connected and disconnected
ones, the latter having a region of instability between the nucleonic
and hybrid configurations. In the following, we focus on the
disconnected topologies.

\section{Mass-radius constraint for hybrid stars}
\label{sec:Models}
\begin{figure*}[tb]
\centering
\ifpdf
\includegraphics[width = 0.92\textwidth]{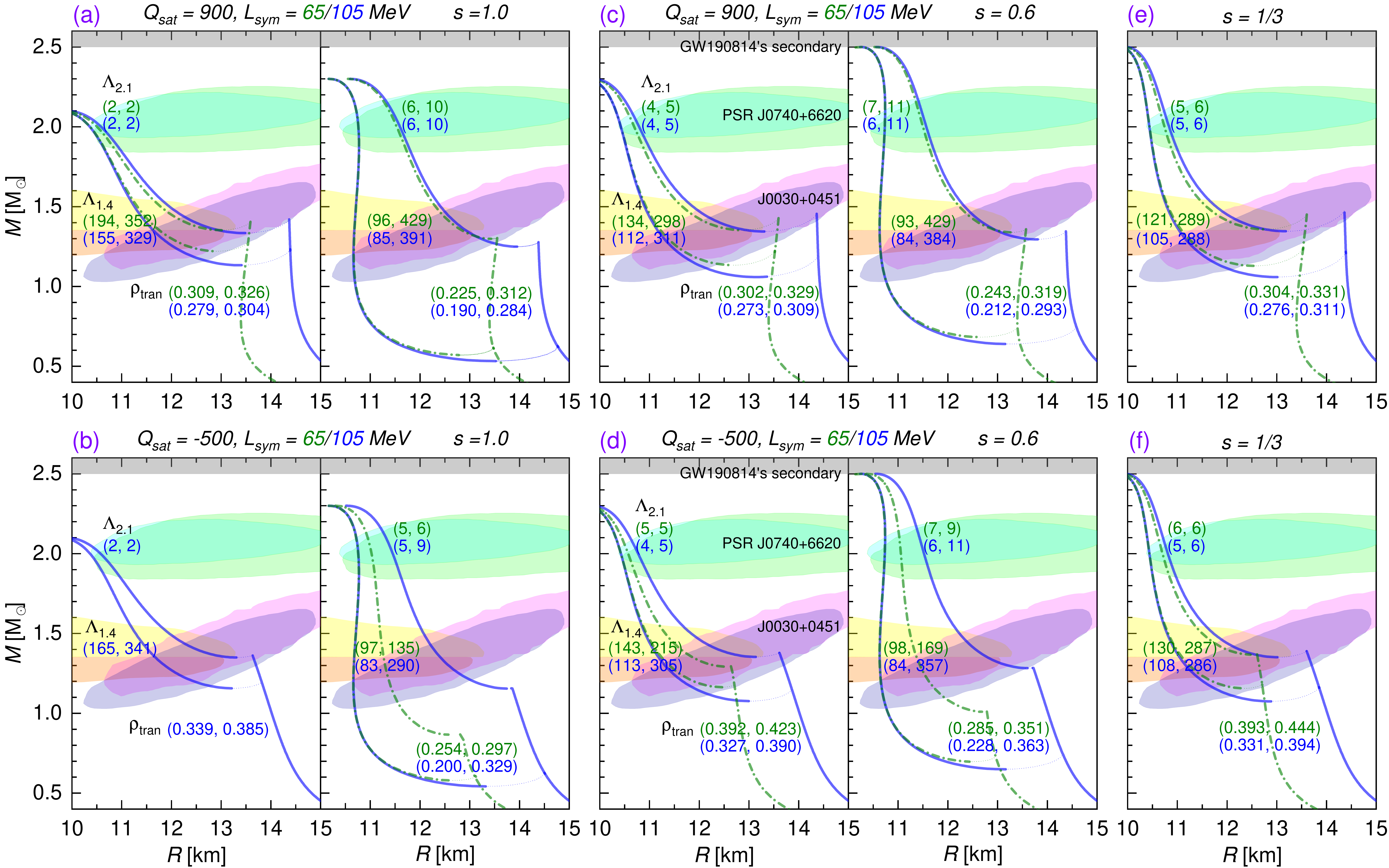}
\else
\includegraphics[width = 0.92\textwidth]{MR_QLS.eps}
\fi
\caption{
Constraints on the \MR relation of CSs featuring twin configurations 
(in some cases, the instability region is not resolved on the figure's scale). Panels (a) and (b): \MR curves for 
hybrid stars characterized by different maximum 
masses (attained on the quark branch) for 
nucleonic EOSs with pairs of values of $Q_{\rm sat}$ (-500 and 900~MeV) and $L_{\rm sym}$ (65 and 105~MeV), and 
quark-matter CSS EOSs with $s=1.0$ (in natural units). The ranges of
transition density $\rho_{\rm tran}$ {in units of fm$^{-3}$}, and the
TDs $\Lambda_{1.4}$ and $\Lambda_{2.1}$ 
(dimensionless) for a canonical mass
$M/M_{\odot} = 1.4$ and massive $M/M_{\odot} 
= 2.1$ stars, respectively, are 
quoted as well. The color coding of 
numbers matches that of the curves. Panels (c), (d) and (e), (f):
same as in (a), (b), but with $s = 0.6$ and $s=1/3$, respectively.}
\label{fig:MR_QLS}
\end{figure*}

To study the occurrence of twin configurations, we select EOSs
from the parameter space of our model that have twin configurations 
and are consistent with both NICER and gravitational wave measurements. Specifically, they yield radii that are above the NICER 90\%-confidence lower limit from PSR J0740+6620 and J0030+0451, 
and below the 90\%-confidence 
upper limit on the radius of a 1.36$M_\odot$ 
  star from the GW170817 event; (the stellar mass  
  is chosen to be the one inferred 
  for an equal-mass binary.)
Figure~\ref{fig:MR_QLS} shows examples of our 
exploration of this parameter space. Each of the six panels in 
Fig.~\ref{fig:MR_QLS} has curves for two nuclear EOS (with the values of 
$\Lsym$ and $\Qsat$ given in each panel title) and a fixed speed of sound 
in the quark-matter EOS (also given in each panel title). Within each 
panel we vary the remaining parameters of the quark-matter EoS (nuclear
matter density at the transition $\rhotran$ and energy density jump
at the transition $\Delta\ep$) as follows: In each of the one or
two plots in each panel, we require the maximum mass on the hybrid
branch $M^{\rm Q}_{\rm max}$ to have a different value (which can
be read off from the $y$ axis), and in each plot, we show results
for two values of $M^{\rm N}_{\rm max}$ which span the range within 
which the \MR curves obey the mass-radius constraints as described 
in the previous paragraph, and contain twin stars (i.e., a 
discontinuity between the nuclear branch and the hybrid branch). 

Let us first focus on the case $s=1.0$ [Figs.~\ref{fig:MR_QLS}(a) and \ref{fig:MR_QLS}(b)]. 
The following systematics are observed {in our setup:} \\
(i). The stiffer is the nucleonic EoS the larger is the range of
masses and radii where twin stars exist; see also Fig.~\ref{fig:MR_Twin} below. Intuitively, this makes sense, as the stiffer nuclear EoS pushes the nucleonic  branch to higher radii, further from the hybrid branch.\\
(ii). The larger the value $M^{\rm Q}_{\rm max}$ the narrower the
range of masses where twins exist and the instability region between
nucleonic and hybrid stars. A higher value, in the current set-up,
requires a smaller $\Delta\ep$ to allow for a steep increase of the
mass. Since for our parameter choice the \MR curves pass through
the GW170817 90\% credible region, the quoted values of TD are consistent
with the upper limit set by the GW170817 on this quantity
event~\citep{LIGO_Virgo_2019}.

Consider next the case of less stiff quark EOS with $s = 0.6$
(Figs.~\ref{fig:MR_QLS}(c) and \ref{fig:MR_QLS}(d))
and $s=1/3$ (Fig.~\ref{fig:MR_QLS}(e) and \ref{fig:MR_QLS}(f)).
The general features found
above are replicated in this case as well. However, in this case
we find twin solutions for models supporting heavier stars,
with $M^{\rm Q}_{\rm max}/M_\odot \gtrsim 2.3$. Interestingly, the
$M^{\rm Q}_{\rm max}/M_{\odot} = 2.5$ sequence, in which the light
component in the GW190814 would be a hybrid star (see
Fig.~\ref{fig:MR_QLS}), allows 
for twin solutions for $s = 0.6,\, 1/3$ in the considered 
range $65\le L_{\rm sym} \le 105$~MeV. 
Compared to the $s = 1.0$ case for the same maximum mass
$M^{\rm Q}_{\rm max}$ the value of $M^{\rm N}_{\rm max}$ is shifted
upwards significantly, which promotes the appearance of twin configurations.

The analysis above shows, among other things, that NICER's 90\% CI
for PSR J0740+6620 does not preclude a strong first-order phase
transition, which (very robustly) leads to a shift of the radius toward smaller values compared to the case without phase transition. Furthermore, 
the  fact that PREX-II analysis of Ref.~\cite{Reed_2021} indicates
$L_{\rm sym} > 65$~MeV at a $68\%$ confidence level
and $L_{\rm sym} > 45$~MeV at $90\%$ corroborates 
the scenario of such a strong
first-order phase transition, 
as a large $L_{\rm sym}$ allows for a
large radius of the nucleonic low-mass stars and thus ``leaves
room'' on the \MR diagram for a hybrid branch with a smaller
radius that is consistent with astrophysical measurements. 
In such a scenario, the appearance of twin configurations is possible. 
This leads us to the conclusion that even the relatively low-mass
$M/M_{\odot}\le 1.5$ CSs could be, in fact, hybrid stars. 

\begin{figure}[tbh]
\centering
\ifpdf
\includegraphics[width = 0.45\textwidth]{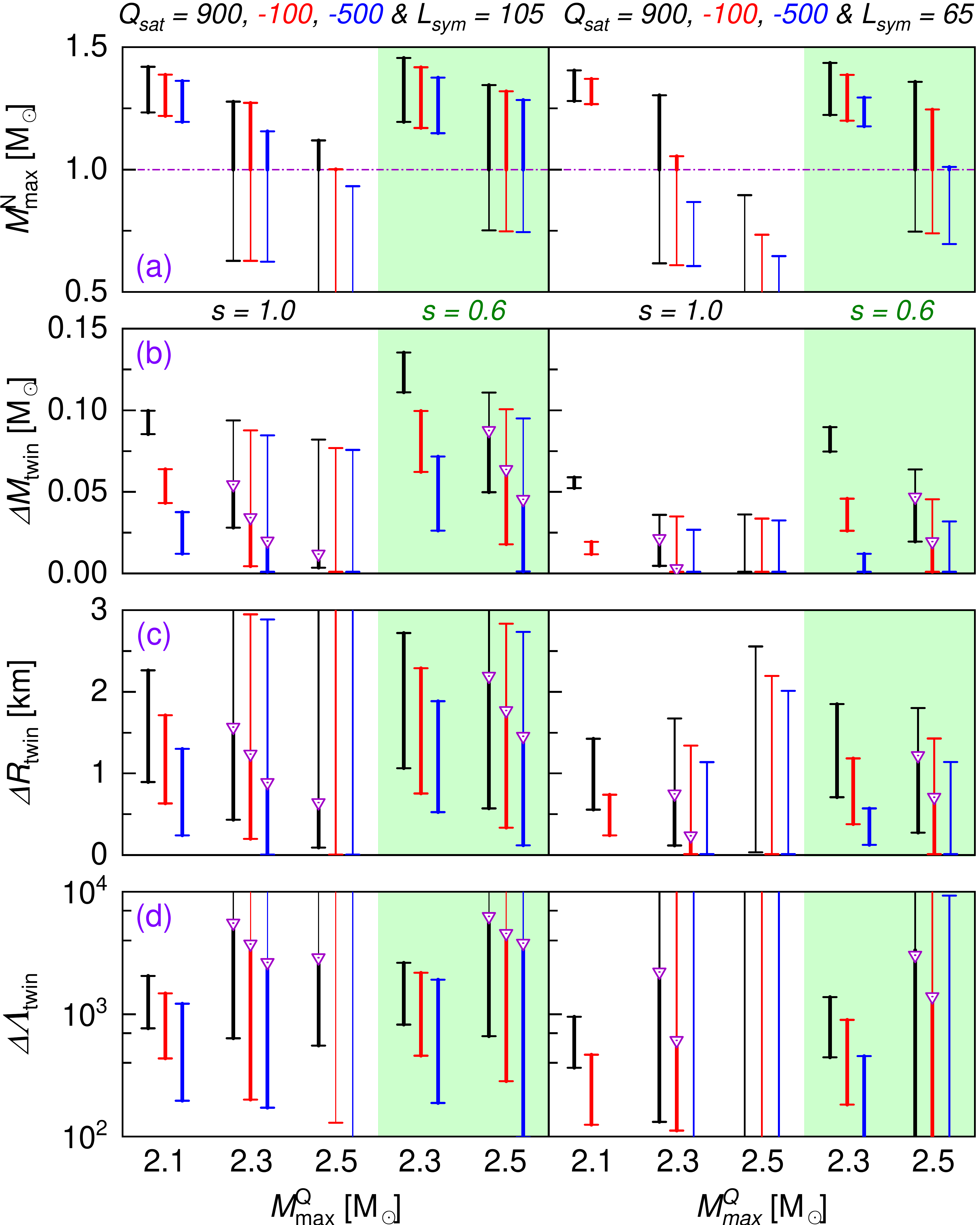}
\else
\includegraphics[width = 0.45\textwidth]{MR_Twin.eps}
\fi
\caption{Ranges of parameters $M^{\rm N}_{\rm max}$, $\Delta M_{\rm twin}$, $\Delta R_{\rm twin}$, and $\Delta \Lambda_{\rm twin}$ characterizing 
twin configurations (see text for definitions) for EOSs constructed 
using nucleonic matter with pairs of values of $Q_{\rm sat}$ 
(-500, -100 and 900~MeV) and $L_{\rm sym}$ (65 and 105~MeV) and 
quark matter with $s =1.0$ and 0.6 (shadowed). 
The triangles followed by the thin lines correspond to twins with 
$M \le 1.0\,M_{\odot}$. See text for discussion.}
\label{fig:MR_Twin}
\end{figure}

Figure~\ref{fig:MR_Twin} shows the values of $M^{\rm N}_{\rm max}$ for 
which twins arise; the mass range of twins $\Delta M_{\rm twin}$, defined as the range between $M^{\rm N}_{\rm max}$ and the minimum value of a CS mass on the hybrid branch; and the intervals 
of the radius $\Delta R_{\rm twin}$ and TD $\Delta\Lambda_{\rm twin}$  associated with this mass range, as 
functions of the parameters defining the EOS. For $s=1.0$ twin configurations
arise for $M^{\rm Q}_{\rm max} /M_\odot \lesssim 2.3$ (within the parameter  
space explored and excluding twins with $M < 1.0\,M_{\odot}$; see below), 
but for $s=0.6$, the values $M^{\rm Q}_{\rm max} /M_\odot \sim 2.5$ 
are obtained. This has implications for the GW190814 event, which involved 
a merger of a 23 $M_{\odot}$ black hole with a $\sim 2.6\,M_\odot$ object. 
Therefore, this object could have been a hybrid star living on a CS 
sequence which contains a twin---i.e., the GW190814 event does not exclude 
twins. (For a contrary view see~\citep{Christian_2021}.)  

In Fig.~\ref{fig:MR_Twin} we see that the range of masses containing twins
is small, $\Delta M_{\rm twin} \lesssim 0.05\,M_\odot$. For radii, note that
the triangles followed by thin lines denote twins with $M<1.0\,M_{\odot}$. 
This is a rough threshold below which stars are unlikely to be observed 
because of their likely instability at the protoneutron-star 
stage~\citep{Gondek_1997,Strobel_2001}. With that condition, the radius difference of twins is $\Delta R_{\rm twin} \sim 1$~km for $s=1.0$; 
the difference is larger for $s=0.6$ with $\Delta R_{\rm twin}$ reaching up 
to 3\,km. This range is largest for very stiff nucleonic EOSs which predict a 
large radius. The narrow range of masses and the modest radius difference
for twins is a consequence of the constraint imposed by the large radius 
of PSR J0740+662, which limits the allowed range of the reduction of the 
radius of a hybrid star and makes it challenging to distinguish between 
nucleonic and hybrid stars as $\Delta R_{\rm twin} = 1$~km requires radius 
measurement accuracy of less than 10\%. However, the difference in the TDs 
of twins can be several hundred to two thousand---see Fig.~\ref{fig:MR_Twin}(d), 
therefore, studies of $\Lambda_1$-$\Lambda_2$  extracted from inspiral phase 
of CS mergers are a more promising avenue for identifying twins.

\section{Conclusions}
\label{sec:Conclusions}

To summarize, we investigated the impact of two recent observational/experimental
results, the inference of the radius of PSR J0740+6620~\cite{NICER_2021a,NICER_2021b} from the x-ray 
data provided by the NICER experiment and
that of the neutron skin thickness by 
the PREX-II experiment in conjunction 
with the DF analysis of  
 Refs.~\cite{Reed_2021,Reinhard_2021}, on the static
properties of relativistic hybrid stars. The analysis of Ref.~\cite{Reed_2021} gives a value of $L_{\rm sym}$, which is in
potential tension with the TD and radius limits inferred from the GW170817 analysis. The $L_{\rm sym}$ 
value extracted in the analysis of Ref.~\cite{Reinhard_2021} is within the accepted range and does not pose any tension with 
multimessanger astrophysics.
The above mentioned new results were confronted here with 
the conjecture of strong first-order phase
transition and the formation of hybrid stars. 
In doing so, we adopted a
density-functional model of nucleonic matter which allows for the
accurate description of low-density nuclear phenomenology, along
with a flexible parametrization of high-density quark matter. 
Our main finding is that it is possible to account for current
astrophysical and nuclear experimental constraints within the scenario
of hybrid stars, which can appear as  twin configurations, 
{\it with a low value of phase transition density from nucleonic to 
quark matter.}
Specifically, in this scenario, the first-order phase transition
{\it naturally softens the EOS} at intermediate densities 
(which is required to avoid the potential tension
between the GW170817 event 
and the PREX-II measurement's interpretation of Ref.~\cite{Reed_2021}), and the assumption of a high sound speed 
in quark matter {\it stiffens the EOS at high densities}, which 
is required to account for $M \gtrsim 2.0\,M_\odot$ massive CS and 
the large radius of PSR J0740+6620 (as measured by the NICER analysis 
teams). We have 
quantified under which conditions such models may have twin solutions. 
In particular, for the squared sound speed in quark matter $s = 0.6$ and 
$s=1/3$, we find twin stars {\em and} a large enough maximum mass to 
allow for a CS in the GW190814 merger event. For low values of $s$, the range of the difference in the radii for twins expands; for example, 
$\Delta R_{\rm twin} \sim 1.0$~km found for $s=1$ becomes 
3~km for $s=0.6$. In the EoS models where
$\Delta$ resonances appear (see Refs.~\citep{Lijj_2020a,Sedrakian_2021} and references therein), the EOS 
softens at intermediate densities, and the radius of the star is reduced with consequences for the \MR relation similar to those found here.

\section*{Acknowledgments}
M.~A. is supported by the U.S. Department of Energy, Office of
Science, Office of Nuclear Physics under Award No. DE-FG02-05ER41375.
J.~L. is supported by the ``Fundamental Research Funds for 
the Central Universities" under Grant No. SWU-020021.
A.~S. is supported by the Deutsche Forschungsgemeinschaft (Grant
No. SE 1836/5-1) and European COST Action ``PHAROS" (CA16214).

\bibliography{Twinstars_refs}

\end{document}